\documentclass[journal]{IEEEtran}

\usepackage{mathtools}
\usepackage{amsmath} 
\usepackage{amssymb}
\usepackage{amsfonts}
\usepackage{verbatim} 
\usepackage{bm,color,soul}
\usepackage{algorithm}
\usepackage{algorithmic} 
\usepackage{cite}
\usepackage{caption2}
\usepackage{stfloats} 
\usepackage[colorlinks, linkcolor=red, anchorcolor=blue, citecolor=green]{hyperref} 

\usepackage{algorithm}
\usepackage{algorithmic}
\usepackage{setspace}
\makeatletter
\newcommand{\removelatexerror}{\let\@latex@error\@gobble}
\makeatother

\usepackage{graphics}
\usepackage{subfigure}  
\usepackage{float}

\usepackage{makecell} 

\ifCLASSINFOpdf
\else
\fi

\usepackage{array}
\usepackage{url}
\usepackage{balance}

\IEEEoverridecommandlockouts

\begin{document}
%
\title{Optimized Joint Beamforming for Wireless Powered Over-the-Air Computation}

 \author{
    \IEEEauthorblockN{ Siyao Zhang, 
        {Xinmin Li},
        {Yin Long},
        {Jie Xu},
	   {and Shuguang Cui}
        }

\thanks{S. Zhang and Y. Long are with the School of Computer Science and Technology, Southwest
University of Science and Technology, Mianyang 621000, China (e-mail: zsy@mails.swust.edu.cn, yinlong@swust.edu.cn).
}
\thanks{X. Li is with the School of Information Engineering, Southwest University of Science and Technology, Mianyang 621000, China (e-mail: lixm@swust.edu.cn).
}
\thanks{J. Xu is with the School of Science and Engineering (SSE) and the Future Network of Intelligence Institute (FNii), The Chinese University of Hong Kong (Shenzhen), Shenzhen 518172, China (e-mail: xujie@cuhk.edu.cn).
}
\thanks{S. Cui is with the SSE and FNii, The Chinese University of Hong Kong (Shenzhen), Shenzhen 518172, China. He is also with Peng Cheng Laboratory (e-mail: shuguangcui@cuhk.edu.cn).
}

\thanks{\emph{Corresponding authors:  Xinmin Li and Yin Long.}}


}

\maketitle

\begin{abstract}
This correspondence studies the wireless powered over-the-air computation (AirComp) for achieving sustainable wireless data aggregation (WDA) by integrating  AirComp and wireless power transfer (WPT) into a joint design. In particular, we consider that a multi-antenna hybrid access point (HAP) employs the transmit energy beamforming to charge multiple single-antenna low-power wireless devices (WDs) in the downlink, and the WDs use the harvested energy to simultaneously send their messages to the HAP for AirComp in the uplink. Under this setup, we minimize the computation mean square error (MSE), by jointly optimizing the transmit energy beamforming and the receive AirComp beamforming at the HAP, as well as the transmit power at the WDs, subject to the maximum transmit power constraint at the HAP and the wireless energy harvesting constraints at individual WDs. To tackle the non-convex computation MSE minimization problem, we present an efficient algorithm to find a converged high-quality solution by using the alternating optimization technique. Numerical results show that the proposed joint WPT-AirComp approach significantly reduces the computation MSE, as compared to other benchmark schemes.
\end{abstract}

%
\IEEEpeerreviewmaketitle
\section{Introduction}
Future wireless networks need to support ubiquitous sensing, communication, and computation of  massive low-power wireless devices (WDs) to enable intelligent Internet-of-things (IoT) applications \cite{zhu2023pushing}. It is thus becoming increasingly important to efficiently aggregate distributed data from these WDs and to provide sustainable energy supply for them. Towards this end, wireless powered over-the-air computation (AirComp) has emerged as a promising solution by integrating AirComp and  wireless power transfer (WPT) into a joint design. In this technique, AirComp enables multiple WDs to simultaneously send their individual data for fast wireless data aggregation (WDA) over the air \cite{2021_ZGX_XJ_WDA_AirComp}, and WPT utilizes radio signals as energy carriers to provide sustainable wireless energy supply for WDs \cite{2019_wireless_powered_commun, 2017_Wireless_Power_Trans}.

Wireless powered AirComp systems, however, face new technical challenges due to the involvement of both AirComp and WPT. First, due to the severe signal propagation loss over distance, far-apart WDs would harvest less energy in the downlink but need more transmit power for AirComp in the uplink, thus inducing the so-called double near-far problem \cite{2013_nearfar} that may significantly degrade the AirComp performance. Next, new energy harvesting constraints are imposed at individual WDs, such that the transmission energy consumption for AirComp at each WD cannot exceed its harvested wireless energy  from the hybrid access point (HAP). How to jointly manage the wireless resource allocation for both downlink WPT and uplink AirComp subject to such constraints is a challenging task for optimizing the AirComp performance.

In the literature, there have been extensive prior works investigating the wireless resource allocations for AirComp (e.g., \cite{2021_ZGX_XJ_WDA_AirComp,2020_CXW_Optimalcontrol_AirComp,2019_ZGX_HKB_MIMO_Sensing, 2021Hybrid,2022_UAV,2022CXW_Federated_Learning}) and WPT (e.g., \cite{2019_wireless_powered_commun,2017_Wireless_Power_Trans}) separately. For instance, the authors in \cite{2020_CXW_Optimalcontrol_AirComp} investigated the transmit power control for minimizing the average computation mean square error (MSE) of AirComp in fading single-input single-output (SISO) channels, and the authors in \cite{2019_ZGX_HKB_MIMO_Sensing, 2021Hybrid} exploited the spatial multiplexing and array gains of multiple-input multiple-output (MIMO) for AirComp via joint transmit and receive beamforming. Then, these designs were extended to unmanned aerial vehicle (UAV)-aided AirComp \cite{2022_UAV} and over-the-air federated edge learning \cite{2022CXW_Federated_Learning}. On the other hand, transmit energy beamforming and waveform optimization have been widely exploited in the WPT literature for enhancing the energy transmission\; efficiency \cite{2019_wireless_powered_commun,2017_Wireless_Power_Trans}. These designs also motivated the applications of simultaneous wireless information and power transfer \cite{2019_wireless_powered_commun}, wireless powered communications \cite{2013_nearfar} and wireless powered task offloading and computing \cite{2018_Joint_Offloading}. By contrast, there have been only a handful of prior works studying the wireless powered AirComp \cite{2019_LXY_HKB_WDA_AirComp,2021_QZB_ZY_AirComp_IRS} under different setups. In particular, the authors in \cite{2019_LXY_HKB_WDA_AirComp} focused on a multi-antenna setup by employing the heuristic time-division energy beamforming (i.e., only one WD is charged at a time)  for WPT and the channel inversion power control for AirComp. Such designs, however, are highly suboptimal in optimizing the AirComp performance in terms of minimizing the computation MSE.



In this correspondence, we consider a multiple-input single-output (MISO) wireless powered AirComp system consisting of a multi-antenna HAP and multiple single-antenna low-power WDs. Under this setup, we jointly exploit the transmit energy beamforming and the receive AirComp beamforming at the HAP as well as the transmit power control at the WDs to resolve the double near-far problem for enhancing the AirComp performance. In particular, our objective is to minimize the computation MSE, by optimizing the joint beamforming at the HAP and the transmit power at the WDs, subject to the transmit power constraint at the HAP and the energy harvesting constraints at individual WDs. Due to the coupling of the transmit power and receive beamforming for AirComp, the formulated computation MSE minimization problem is highly non-convex and difficult to solve. To tackle this issue, we present an efficient algorithm to find a converged high-quality solution to this problem by using the alternating optimization technique, in which the transmit energy beamforming (together with WDs' power control) and the receive beamforming are alternately optimized. Numerical results show that the proposed joint WPT-AirComp design significantly reduces the computation MSE, as compared to the benchmark scheme in \cite{2019_LXY_HKB_WDA_AirComp} and the design with isotropic energy transmission, especially when the WDs are located at different distances with the HAP.
\section{System Model}
We consider a MISO wireless powered AirComp system, which consists of one HAP equipped with $M \ge$ 1 antennas, and a set $\mathcal{K}$ $\triangleq$ $\lbrace1,...,K\rbrace$ of WDs each equipped with one antenna. We consider the time-division duplex (TDD) transmission protocol, in which the transmission block is divided into two time slots for downlink WPT and uplink AirComp, respectively.  Suppose that for a given transmission block with unit duration, the durations for WPT and AirComp are given by $\alpha_1$ and $\alpha_2$, respectively, with $\alpha_1 + \alpha_2 = 1$.  

We consider quasi-static channel models, in which the channels remain unchanged over the transmission block of our interest but may change over different blocks. Let $\boldsymbol h_k \in \mathbb{C}^{M \times 1}$ denote the channel vector between the HAP and each WD $k\in\mathcal{K}$. It is assumed that the HAP has the perfect channel state information (CSI) of channel vector $\boldsymbol h_k$ and each WD $k$ has its own CSI of $\boldsymbol h_k$, which can be obtained based on the reverse-link channel estimation by exploiting the uplink-downlink channel reciprocity.

First, we consider the WPT from the HAP to the WDs in the downlink. Let $\boldsymbol x \in \mathbb{C} ^{M\times 1} $ denote the transmit energy signal by the HAP, which is a randomly generated sequence with $\boldsymbol S = \mathbb{E}[\boldsymbol x \boldsymbol x^{H}]\succeq \bm{0}$ denoting the transmit covariance matrix. Suppose that the HAP is subject to a maximum transmit power budget $P$. We thus have $\mathbb{E}[||\boldsymbol x|| ^2 ] \!=\! \mathrm{tr} (\boldsymbol S) \!\leq P$. By considering a linear energy harvesting model \cite{2019_wireless_powered_commun}, the  harvested wireless power at WD $k$ is
\begin{align}\label{eq1}
  E_k = \eta \mathbb{E}[|\boldsymbol  h_k^H \boldsymbol x|^2]=\eta \boldsymbol h_k^H\boldsymbol S\boldsymbol h_k,
\end{align}
where $0 < \eta < 1$ denotes the energy harvesting efficiency at each WD.

	Next, we consider the AirComp from the WDs to the HAP in the uplink. Let $s_k$ denote the transmit signal at WD $k$, where $s_k$'s are independent random variables with zero mean and unit variance. Suppose that the HAP is interested in computing the mean value of $s_k$'s, given by
\begin{align}
  f=\frac{1}{K}\sum\nolimits_{k\in\mathcal{K}}s_k.
\end{align}
For each WD $k$, let $b_k$ denote the transmit coefficient and $b_ks_k$ denote the corresponding transmit signal. Accordingly, the transmit power at each WD $k\in\mathcal K$ is $\mathbb{E}[|b_ks_k|^2] = |b_k|^2$. As such, the received signal at the HAP is given by 
\begin{align}
\boldsymbol r = \sum \nolimits_{k\in\mathcal K} \boldsymbol h_k b_k s_k + \boldsymbol z, 
\end{align}
where $\boldsymbol z$ denotes the additive white Gaussian noise (AWGN) at the HAP receiver with zero mean and variance $\sigma^2$. After receiving $\boldsymbol r$, the HAP implements the receive beamforming vector $\boldsymbol w$ and multiplies it by $\frac{1}{K}$. As a result, the processed signal is given as $\hat f$ in the following, which is used as the estimate of $f$ of our interest.
\begin{align}
  \hat{f}=\frac{1}{K}\boldsymbol w^H \boldsymbol r=\frac{1}{K}\boldsymbol w^H(\sum\nolimits_{k\in\mathcal {K}}\boldsymbol h_k b_k s_k + \boldsymbol z)
\end{align}
Notice that the transmission energy consumption at each WD $k$ over duration $\alpha_2$ cannot exceed the wireless harvested from the HAP over duration $\alpha_1$. Therefore, we have the following energy harvesting constraints: 
\begin{align}\label{eq5}
  \alpha_2 |b_k|^2 \leq \alpha_1 E_k = \alpha_1\eta \boldsymbol h_k^H\boldsymbol S\boldsymbol h_k,\forall k \in \mathcal{K}.
\end{align}

We are interested in minimizing the distortion of the recovered average $\hat f$ with respect to the ground truth $f$, which is measured by the MSE given by
\begin{align}\label{eq6}
  \!\text{MSE}\!&=\!\mathbb{E}[(\hat{f}\!-\!f)^2]\!=\!\frac{1}{K^2}\mathbb{E}[(\sum\nolimits_{k\in\mathcal {K}}\!s_k(\boldsymbol w^H\boldsymbol h_k b_k\!-\!1)\!+\!\boldsymbol w^H \boldsymbol z)^2]\nonumber\\ &= \frac{1}{K^2}(\sum\nolimits_{k\in\mathcal {K}}|\boldsymbol w^H\boldsymbol h_k b_k - 1|^2 + ||\boldsymbol w||^2\sigma^2),
\end{align}
where the expectation is taken over the randomness of both ${\{s_k\}}$ and $\boldsymbol z$. 

Our objective is to minimize the MSE in (\ref{eq6}), by jointly optimizing the transmit power or coefficient $\{b_k\}$ at the WDs as well as the receive AirComp beamforming vector $\boldsymbol w$ and the transmit beamforming or covariance $\boldsymbol S$ at the HAP, subject to the energy harvesting constraints in (\ref{eq5}) and the transmit power constraint at the HAP. Therefore, the computation MSE minimization problem is formulated as
\begin{subequations}
\begin{align}
\text{(P1)}:\!\!\min_{\mbox{\tiny$\begin{array}{c}
 \{b_k\},\boldsymbol  w, \boldsymbol S\succeq \boldsymbol{0} \ 
\end{array}$}}
&\sum\nolimits_{k \in \mathcal K} |\boldsymbol w^H \boldsymbol h_k b_k - 1|^2 + ||\boldsymbol w||^2 \sigma^2\label{eq7a}\\
\text{s.t.}\quad &\alpha_2 |b_k|^2 \leq \alpha_1\eta \boldsymbol h_k^H\boldsymbol S\boldsymbol h_k, \forall k \in \mathcal{K}\label{eq7b}\\
&\mathrm{tr}(\boldsymbol S) \leq P\label{eq7c}.
\end{align}
\end{subequations}
It is observed that the objective function in (P1) consists of two terms, namely the signal misalignment error (i.e., $\sum_{k \in \mathcal K} |\boldsymbol w^H \boldsymbol h_k b_k - 1|^2$) and the noise-induced error (i.e., $||\boldsymbol w||^2 \sigma^2$). Therefore, minimizing the MSE needs to properly balance the tradeoff between the two terms. Furthermore, in order to avoid the double near-far problem in the wireless powered AirComp system, the HAP needs to properly adjust both $\boldsymbol S$ and $\boldsymbol w$ to steer the transmit and receive powers towards the far-apart WDs to optimize the AirComp performance. Nevertheless, problem (P1) is highly non-convex due to the coupling of the denosing factor $\boldsymbol  w$ and the transmit power $\{b_k\}$. Therefore, it is difficult to be optimally solved.
\section{Proposed Joint Beamforming Solution}
In this section, we present an alternating optimization based algorithm to find a converged and high-quality solution to problem (P1), in which $\{b_k\}$ / $\boldsymbol S$ and $\boldsymbol w$ are optimized in an alternating manner by considering the other to be given. 

First, we optimize the transmit power $\{b_k\}$ at the WDs and the transmit energy beamformers $\boldsymbol S$ at the HAP with any given $\boldsymbol w$. Without loss of optimality, we set
\begin{align}
b_k = \frac{(\boldsymbol w^H \boldsymbol h_k)^H}{|\boldsymbol w^H \boldsymbol h_k|}\tilde b_k, \forall k\in \mathcal K,\label{eq8}
\end{align}
such that the MSE objective in (\ref{eq7a}) is minimized with phase alignment, where $\tilde b_k \geq 0$ denotes the transmit amplitude of WD $k \in \mathcal K$. By substituting this into (P1) and omitting the constant term $||\boldsymbol  w||^2\sigma^2$ in the objective function, the optimization of $\{\tilde b_k\}$ and $\boldsymbol S$ is expressed as
\begin{subequations}
\begin{align}
\text{(P2)}:\min_{\mbox{\tiny$\begin{array}{c}
 \{\tilde b_k\geq 0\}, \boldsymbol S \succeq \boldsymbol{0}\ 
\end{array}$}}
&\sum\nolimits_{k \in \mathcal K} (|\boldsymbol w^H \boldsymbol h_k| \tilde b_k - 1)^2 \label{eq9a}\\
\text{s.t.}\quad\ &\alpha_2 \tilde b_k^2 \leq \alpha_1\eta \boldsymbol h_k^H\boldsymbol S\boldsymbol h_k, \forall k \in \mathcal{K}\label{eq9b}\\
&\mathrm{tr}(\boldsymbol S) \leq P.\label{eq9c}
\end{align}
\end{subequations}
Notice that problem (P2) is convex as the objective function in (\ref{eq9a}) is convex and the constraints in (\ref{eq9b}) and (\ref{eq9c}) are also convex. Furthermore, problem (P2) satisfies the Slater's conditions, and thus the strong duality holds between (P2) and its dual problem. Therefore, problem (P2) can be solved optimally by using Lagrange-duality method. Let \!$\{\mu_k\}$ and $\nu$ denote the non-negative Lagrange multipliers associated with the constraints in (\ref{eq9b}) and (\ref{eq9c}) in problem (P2), respectively. The Lagrangian is
\begin{align}
\mathcal L(\{\tilde b_k\},&\boldsymbol S,\{\mu_k\},\nu) = \sum_{k \in \mathcal K}\left(\left(|\boldsymbol w^H \boldsymbol h_k| \tilde b_k - 1\right)^2 + \mu_k \alpha_2 \tilde b_k^2\right)\nonumber\\  
 &+ \mathrm{tr}\left(\left(\nu\boldsymbol I - \sum_{k \in \mathcal K}\alpha_1\eta \mu_k \boldsymbol h_k\boldsymbol h_k^H\right)\boldsymbol S\right) - \nu  P,\label{eq10}
\end{align}
and the corresponding dual function is
\begin{align}
g(\{\mu_k\},\nu) = \min_{\{\tilde b_k\geq 0\}, \boldsymbol S \succeq \boldsymbol{0}}\mathcal L(\{\tilde b_k\},\boldsymbol S,\{\mu_k\},\nu).\label{eq11}
\end{align}

\newtheorem{lemma}{Lemma}
\begin{lemma}
In order for the dual function $g(\{\mu_k\},\nu)$ to be lower bounded from below, it follows that $\boldsymbol F(\boldsymbol\mu) \triangleq \nu\boldsymbol I -  \sum_{k \in \mathcal K}\alpha_1\eta \mu_k \boldsymbol h_k\boldsymbol h_k^H \succeq \boldsymbol{0}$.
\end{lemma}

\begin{IEEEproof}[Proof]
	See Appendix \ref{Appendix A}.
\end{IEEEproof}

Accordingly, the dual problem of (P2) is
\begin{subequations}
\begin{align}
\text{(D2)}: \max_{\mbox{\tiny$\begin{array}{c}
 \{\mu_k\geq 0\}, \nu \geq 0\ 
\end{array}$}}
&g(\{\mu_k\},\nu)\label{eq12a}\\
\text{s.t.}\quad\ &\boldsymbol F(\boldsymbol\mu) \succeq \boldsymbol{0}.\label{eq12b}
\end{align}
\end{subequations}

In the following, we solve problem (P2) by equivalently solving dual problem (D2). In particular, we first solve problem (\ref{eq11}) to obtain dual function $g(\{\mu_k\},\nu)$ under any given $\{\mu_k\}$ and $\nu$, and then search over them to solve problem (D2) via the subgradient-based methods such as the ellipsoid method \cite{2019_S.Boyd_ellipsoid_convex_opt}. Let $\{\mu_k^*\}$ and $\nu^*$ denote the optimal dual solution to problem (D2), for which detailed derivation can be found in Appendix \ref{Appendix B}. Then we have the optimal solution to (P2) in the following proposition.  

\newtheorem{proposition}{Proposition}
\begin{proposition}
The optimal solution of transmit amplitude $\{\tilde b_k\}$ to  problem (P2) is given by
\begin{align}
\tilde {b}_k^* = \frac{|\boldsymbol w^H \boldsymbol h_k|}{|\boldsymbol w^H \boldsymbol h_k|^2 + \mu_k^*\alpha_2},\forall k\in \mathcal K. \label{eq13}
\end{align}
Accordingly, the optimal transmit energy covariance $\boldsymbol S^*$ to (P2) is the solution to the following convex feasibility semi-definite program (SDP) that is solvable via standard convex optimization techniques such as CVX \cite{2009_CVX}.
\begin{align}
\text{Find} \;&\boldsymbol S 
\label{eq14}\\
\text{s.t.}\;&\alpha_2 (\tilde b_k^*)^2 \leq \alpha_1\eta \boldsymbol h_k^H\boldsymbol S\boldsymbol h_k, \forall k \in \mathcal{K}\nonumber\\
&\mathrm{tr}(\boldsymbol S) \leq P, \boldsymbol S \succeq \boldsymbol 0.\nonumber
\end{align}
\begin{IEEEproof}[Proof]
	See Appendix \ref{Appendix B}.
\end{IEEEproof}

\end{proposition}

Next, we optimize the receive AirComp beamformer $\boldsymbol w$ with fixed $\{b_k\}$ and $\boldsymbol S$ in problem (P1). Accordingly, the optimization problem becomes
\begin{align}
\text{(P3)}: \min_{\mbox{\tiny$\begin{array}{c}
\boldsymbol w \\
\end{array}$}}
\quad\sum\nolimits_{k \in \mathcal K} |\boldsymbol w^H \boldsymbol h_k b_k - 1|^2 + ||\boldsymbol w||^2 \sigma^2.\label{eq15}
\end{align}

By checking the gradient of the objective function in problem \!(P3), the optimal \!solution\! to problem (P3) is obtained as
\begin{align}
\boldsymbol  w^* = (\sum\nolimits_{k \in \mathcal K}|b_k|^2 \boldsymbol h_k \boldsymbol h_k^H + \sigma^2 \boldsymbol I)^{-1}\sum\nolimits_{k \in \mathcal K} b_k\boldsymbol h_k.\label{eq16}
\end{align}

In summary, the alternating optimization based algorithm for solving (P1) is implemented iteratively as follows. In each iteration, we first obtain the transmit power $\{b_k^*\}$ and the energy beamforming $\boldsymbol S^*$ based on (\ref{eq8}) and Proposition 1, and then update $\boldsymbol w^*$ based on (\ref{eq16}). The iteration terminates when the decrement of MSE is less than a certain threshold. Note that after each iteration, the updated computation MSE objective is monotonically non-increasing. As the MSE objective is lower bounded, the convergence of the algorithm is ensured. 
\newtheorem{remark}{Remark}
\begin{remark}
It is interesting to discuss the obtained solution structure for gaining more insights. First, it is observed from (\ref{eq13}) that the transmit power at each WD $k$ follows the truncated channel inversion power control, in which the dual variable $\{\mu_k^*\}$ serves as the regularization parameter. In paticular, if the WD $k$ has sufficient harvested power or the energy harvesting constraint in (\ref{eq9b}) is inactive, then we have $\mu_k^* = 0$ based on the complementary slackness condition, such that $\tilde b_k = \frac{|\boldsymbol w^{*H} \boldsymbol h_k|}{|\boldsymbol w^{*H} \boldsymbol h_k|^2}$ follows the channel inversion power control. By contrast, if $\mu_k^* > 0$, then the energy harvesting constraints in (\ref{eq9b}) is active, i.e., WD $k$ should use up its harvested power by the truncated channel inversion power control in (\ref{eq13}). Furthermore, if the harvested power at WD $k$ is more insufficient (or equivalently WD $k$ is far apart from the HAP), then $\{\mu_k^*\}$ should be greater. 

Next, we discuss the transmit energy beamformers $\boldsymbol S^*$ in (\ref{eq14}) and the receive AirComp beamformer $\boldsymbol w^*$ in (\ref{eq16}). In particular, $\boldsymbol w^*$ follows a sum minimum MSE (MMSE) structure for aggregating the signals from all the WDs effectively. If WD $k$ is far apart from the HAP or the channel $\boldsymbol h_k$ becomes poor, then it follows from (\ref{eq14}) and (\ref{eq16}) that both $\boldsymbol S^*$ and $\boldsymbol w^*$ should be designed towards these WDs, thus enhancing their performances and resolving the double near-far problem. 
\end{remark}
\begin{figure}[t]
\setlength{\abovecaptionskip}{-3pt}
\setlength{\belowcaptionskip}{0pt}
\centering
\includegraphics[width= 0.48\textwidth]{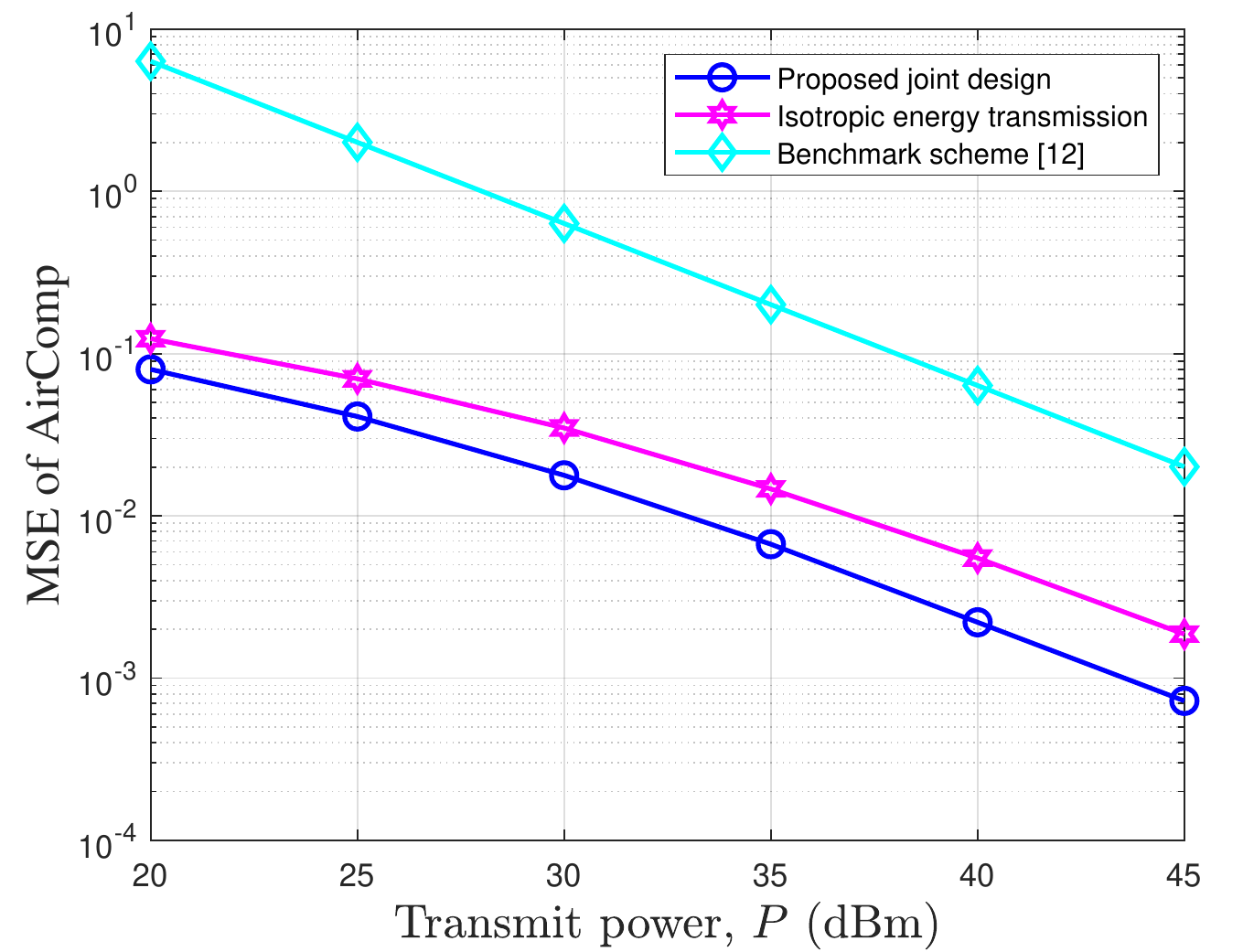} 
~\\
\!                    
\caption{ The computation MSE versus the transmit power when the distances between different WDs and HAP are same.}
\label{figure2}
\end{figure}
\begin{figure}[t]
\setlength{\abovecaptionskip}{-3pt}
\setlength{\belowcaptionskip}{0pt}
\centering
\includegraphics[width= 0.48\textwidth]{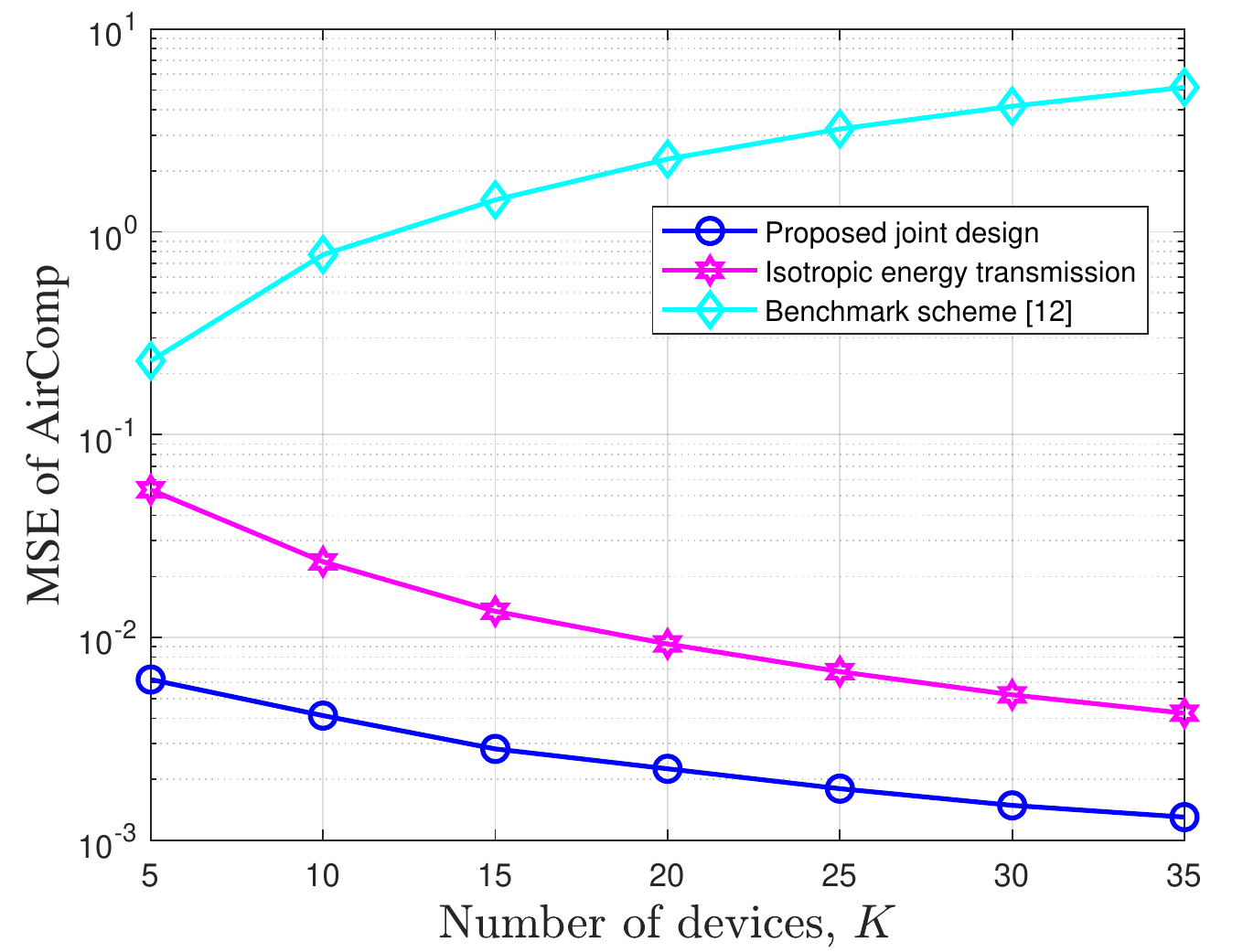}
~\\
\!                    
\caption{ The computation MSE versus the number of WDs.}
\label{figure3}
\end{figure}

\begin{figure}[t]
\setlength{\abovecaptionskip}{-3pt}
\setlength{\belowcaptionskip}{0pt}
\centering
\includegraphics[width= 0.48\textwidth]{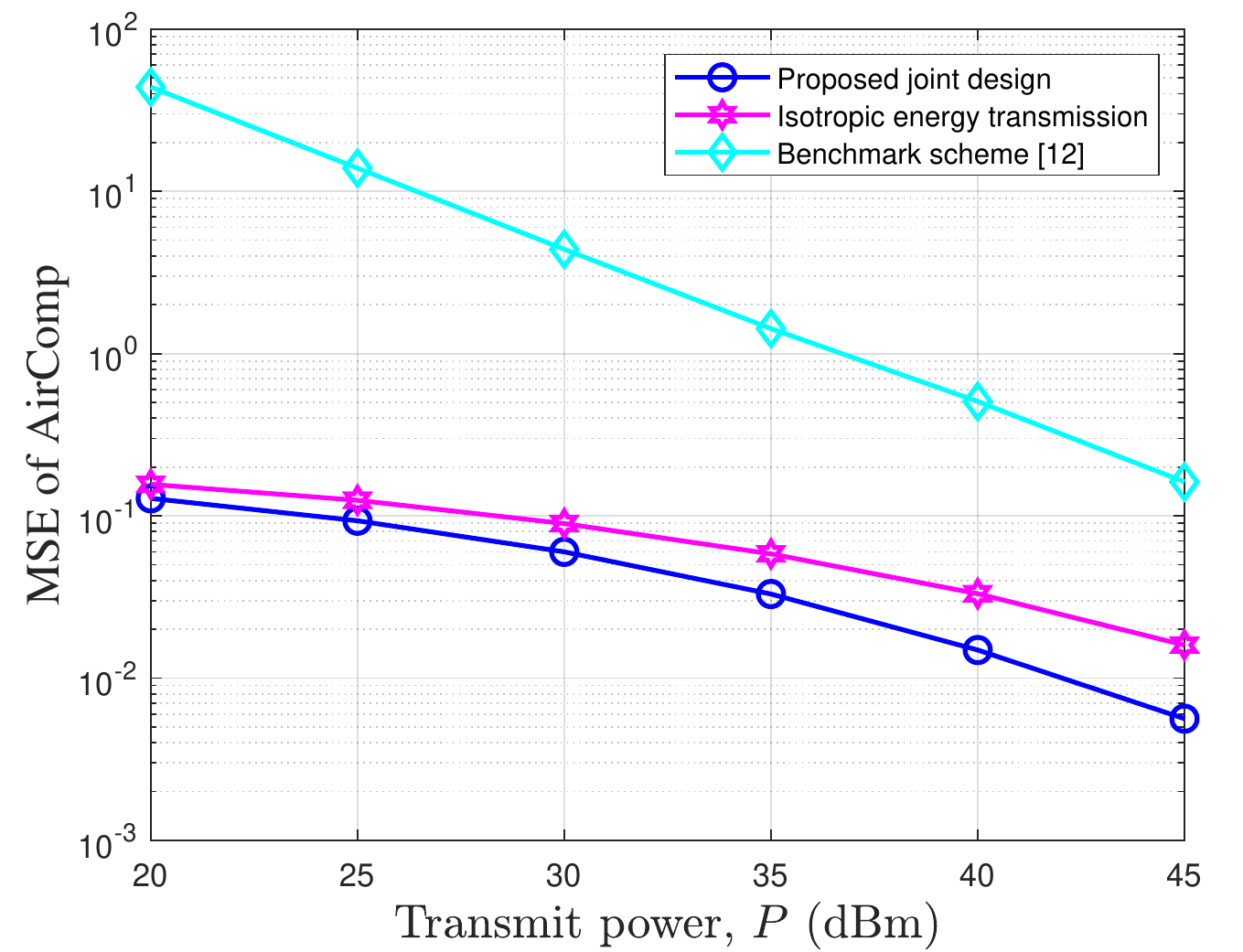} 
~\\
\!                    
\caption{ The computation MSE versus the transmit power when the distances between different WDs and HAP are different.}
\label{figure4}
\end{figure}
\section{Numerical Results}
In this section, we provide numerical results to evaluate the performance of our proposed joint beamforming design. In the simulation, we consider the path loss model, $L(d) = K_0(d/d_0)^{-\alpha_0}$, where $K_0 = -30 $ dB is the pathloss constant at the reference distant $d_0$  = 1 meter (m), $\alpha_0$  = 3 is the path loss exponent, and $d$ is the distance between the HAP and WDs. We also consider Rician fading with the Rician factor being 5. We consider the following two benchmark schemes for performance comparison.

$\emph{1) Benchmark scheme}$ \cite{2019_LXY_HKB_WDA_AirComp}: This scheme employs the time-division energy beamforming for WPT and the channel inversion power control for AirComp. 

$\emph{2) Isotropic energy transmission\\}$: The HAP sets $\boldsymbol S = \{ P/M\} \boldsymbol I$ for WPT. Accordingly, the transmit power $\{b_k\}$ and the receive beamforming $\boldsymbol w$ are designed by using the alternating optimization, similarly as in our proposed design. 

Fig. \ref{figure2} shows the average MSE versus the transmit power $P$ when the WDs are located with same distances of 10 m with the HAP, where $M = 4$, $K = 4$, and $\sigma^2 = -100$ dBm. It is observed that our proposed joint design achieves the lowest computation MSE in the whole regime of the transmit power $P$, and the isotropic energy transmission outperforms the benchmark in \cite{2019_LXY_HKB_WDA_AirComp}. This shows the importance of the joint optimization of the transmit energy and receive beamforming at the HAP as well as the transmit power at the WDs.

Fig. \ref{figure3} shows the computation MSE versus the number of WDs $k$ with $M = 40$, $P = 20 $ dBm, and $\sigma^2 = -100$ dBm, where the distances between the HAP and WDs are 10 m. It is observed that as the value of $K$ increases, the computation MSE values achieved by the proposed joint design and the isotropic energy transmission monotonically decrease. This is due to the fact that with proper joint  beamforming and power control in this case, the HAP can properly aggregate more data for averaging. By contrast, it is observed that when $K$ increases, the computation MSE by the benchmark \cite{2019_LXY_HKB_WDA_AirComp} increases. This is because with highly suboptimal time-division energy beamforming and channel inversion power control in this case, the MSE performance is constrained by the worst-case WD, which increases with increasing $K$. 

Fig. \ref{figure4} shows the computation of MSE versus the transmit power $P$ with $M = 4$, $K = 4$, and $\sigma^2 = -100$ dBm, where the distances between the HAP and the four WDs are 5m, 10m, 15m, 20m, respectively. By comparing it with Fig. \ref{figure2}, it is observed that with different distances, the achieved MSE values by the three schemes increase, especially that by the benchmark scheme \cite{2019_LXY_HKB_WDA_AirComp}, as the performance is limited by the most far-apart WD with the poorest channel condition, due to the double near-far problem. More specifically, the performance gap between the proposed design versus the isotropic energy transmission or the benchmark scheme \cite{2019_LXY_HKB_WDA_AirComp} is observed to be more significant when $P$ becomes large. This shows the benefit of our joint beamforming and power control design in this case. 
\section{Conclusion}
This correspondence studied a new wireless powered AirComp system for achieving sustainable AirComp by integrating with the emerging WPT technique. By considering a MISO setup, we proposed to jointly design the transmit energy and receive AirComp beamforming at the HAP as well as the transmit power control at the WDs, for optimizing the computation MSE, by taking into account the newly introduced energy harvesting constraints. Numerical results showed the performance gain of the proposed design over other benchmarks.

\begin{appendix}
\subsection{Proof of Lemma 1}\label{Appendix A}
We prove $\boldsymbol F(\boldsymbol\mu) \succeq \boldsymbol{0}$ by contradiction. Assume that $\boldsymbol F(\boldsymbol\mu)$ is not positive semidefinite. In this case, there exists an eigenvector  $\boldsymbol \xi$ corresponding to a negative eigenvalue of $\boldsymbol F(\boldsymbol\mu)$. By setting  $\boldsymbol S  = \tau \boldsymbol \xi \boldsymbol \xi^H \succeq \boldsymbol{0}$ with $\tau \rightarrow \infty$, it follows that 
\begin{equation}
\lim_{\tau \to +\infty} \mathrm{tr}(\boldsymbol F(\boldsymbol\mu) \boldsymbol S) = \lim_{\tau \to +\infty} \tau  \boldsymbol \xi^H \boldsymbol F(\boldsymbol\mu) \boldsymbol \xi = -\infty.\label{eq17}
\end{equation}
This shows that the dual function value becomes unbounded from below. This introduces a contradiction. Therefore, it must hold that $\boldsymbol F(\boldsymbol\mu) \succeq \boldsymbol{0}$. This completes the proof of Lemma 1.

\subsection{Proof of Proposition 1}\label{Appendix B}
First, we solve problem (\ref{eq10}) with given $\{\mu_k\}$ and $\nu$, which can be decomposed into the following $(K+1)$ subproblems for optimizing $\boldsymbol S$ and $\{\tilde b_k\}$, respectively.
\begin{align}
\min_{\boldsymbol S\succeq \boldsymbol{0} }\quad
\mathrm{tr}(\boldsymbol F(\boldsymbol\mu)\boldsymbol S)\label{eq18}
\end{align}
\vspace{-5pt}
\begin{align}
\min_{\{\tilde b_k \geq 0\}}\quad(|\boldsymbol w^H \boldsymbol h_k| \tilde b_k - 1)^2 + \mu_k \alpha_2\tilde b_k^2\label{eq19}
\end{align}
For problem (\ref{eq18}), as $\boldsymbol F(\boldsymbol\mu) \succeq \boldsymbol{0}$, the optimal solution can be any positive semi-definite matrix. We simply choose $\boldsymbol S = \boldsymbol 0$ to  find the dual function $g(\{\mu_k\}, \nu)$. For problem (\ref{eq19}), by checking the first derivative, we have the optimal solution as
\begin{align}
\tilde b_k^\ddagger = \frac{|\boldsymbol w^H \boldsymbol h_k|}{|\boldsymbol w^H \boldsymbol h_k|^2 + \mu_k \alpha_2}.\label{eq20}
\end{align}
Therefore, the dual function $g(\{\mu_k\}, \nu)$ is obtained. 

Next, we solve dual problem (D1) by optimizing $\{\mu_k\}$ and $\nu$. As the dual function $g(\{\mu_k\}, \nu)$ is concave but non-differentiable in general, we use subgradient based methods such as the ellipsoid method \cite{2019_S.Boyd_ellipsoid_convex_opt} to find the optimal $\{\mu_k^*\}$ and $\nu^*$. In order to implement the ellipsoid method, we need to find the subgradients of the objective function in (\ref{eq12a}) and the constraint function in (\ref{eq12b}). For the objective function (\ref{eq12a}), one subgradient is
\begin{align}
[\alpha_2 \tilde{b}_1^{\ddagger2} \text{, \dots, } \alpha_2\tilde{b}_K^{\ddagger2}\text{,}-P]^\dagger.\label{eq21}
\end{align} 

We have the following lemma to obtain the  subgradient of the constraint function in (\ref{eq12b}).
\begin{lemma}
Let $\boldsymbol \delta\in \mathbb{C}^{M \times 1}$ denote that the eigenvector corresponding to the smallest eigenvalue of $\boldsymbol F(\boldsymbol\mu)$, i.e., $\boldsymbol \delta = \mathop{\arg\min}_{||\boldsymbol\xi||=1}\boldsymbol \xi^H\boldsymbol F(\boldsymbol\mu)\boldsymbol\xi$. Therefore, the constraint $\boldsymbol F(\boldsymbol\mu) \succeq \boldsymbol{0}$ is equivalent to a new constraint  $\boldsymbol \delta^H\boldsymbol F(\boldsymbol\mu)\boldsymbol\delta \geq 0$, and the subgradient of function $\boldsymbol \delta^H\boldsymbol F(\boldsymbol\mu)\boldsymbol\delta$ is
\begin{align}
[\alpha_1\eta \boldsymbol \delta^H\boldsymbol h_1\boldsymbol h_1^H\boldsymbol \delta \text{, \dots, }\alpha_1\eta \boldsymbol \delta^H\boldsymbol h_K\boldsymbol h_K^H\boldsymbol \delta\text{,}-\boldsymbol\delta^H\boldsymbol\delta]^\dagger.\label{eq22}
\end{align}
\end{lemma}
\begin{IEEEproof}[Proof]
	This lemma can be similarly proved by \cite{2018_Joint_Offloading}, [Appendix D], for which the details are omitted. 
\end{IEEEproof}

Next, with the optimal dual solutions $\{\mu_k^*\}$ and $\nu^*$ at hand, we substitute them in (\ref{eq20}) and accordingly obtain the optimal solution of $\{\tilde b_k\}$ to problem (P2) as $\{\tilde b_k^*\}$ in (\ref{eq13}). Furthermore, it remains to determine the optimal solution of $\boldsymbol S$. By substituting $\{\tilde b_k^*\}$ in (P2), we obtain $\boldsymbol S^*$ by solving the feasibility SDP in (\ref{eq14}).  
This thus completes the proof of Proposition 1. 
\end{appendix}
\bibliographystyle{IEEEtran}
\bibliography{IEEEabrv,WPT_AirComp}


\end{document}